\documentclass[twocolumn,prl]{revtex4-1}
\usepackage{graphicx}
\graphicspath{ {./images/} }
\usepackage{dcolumn}
\usepackage{bm}
\usepackage{siunitx}
\usepackage{hyperref}
\usepackage{color}
\usepackage{ulem}
\usepackage{amsthm,amsmath}
\usepackage[utf8]{inputenc}
\usepackage[mathlines]{lineno}

\usepackage[colorinlistoftodos]{todonotes}
\presetkeys{todonotes}{inline,backgroundcolor=yellow}{}

\begin{document}

\title{Mercury Hydroxide as a Promising Triatomic Molecule to Probe P,T-odd Interactions}

\author{R. Mitra}
\affiliation{Atomic, Molecular and Optical Physics Division, Physical Research Laboratory, Navrangpura, Ahmedabad 380009, India and \\
Indian Institute of Technology Gandhinagar, Palaj, Gandhinagar 382355, Gujarat, India}

\author{V. S. Prasannaa}
\email{srinivasaprasannaa@gmail.com}
\affiliation{Atomic, Molecular and Optical Physics Division, Physical Research Laboratory, Navrangpura, Ahmedabad 380009, India}

\author{B. K. Sahoo}
\email{bijaya@prl.res.in}
\affiliation{Atomic, Molecular and Optical Physics Division, Physical Research Laboratory, Navrangpura, Ahmedabad 380009, India}

\author{X. Tong}
\affiliation{State Key Laboratory of Magnetic Resonance and Atomic and Molecular Physics, Wuhan Institute of Physics and Mathematics, Chinese Academy of Sciences, Wuhan 430071, China}

\author{M. Abe}
\affiliation{Tokyo Metropolitan University, 1-1, Minami-Osawa, Hachioji-city, Tokyo 192-0397, Japan}

\author{B. P. Das}
\affiliation{Department of Physics, Tokyo Institute of Technology,
2-12-1-H86 Ookayama, Meguro-ku, Tokyo 152-8550, Japan}

\date{\today}

\begin{abstract}
In the quest to find a favourable triatomic molecule for detecting electric dipole moment of an electron (eEDM), we identify mercury hydroxide (HgOH) as an extremely attractive candidate from both experimental and theoretical viewpoints. Our calculations show that there is a  four-fold enhancement in the effective electric field of HgOH compared to the recently proposed ytterbium hydroxide (YbOH) [Phys. Rev. Lett. {\bf 119}, 133002 (2017)] for eEDM measurement. Thus, in the (010) bending state associated with the electronic ground state, it could provide better sensitivity than YbOH from a  theoretical point of view. We have also investigated the potential energy curve and permanent electric dipole moment of HgOH, which lends support for its experimental feasibility. Moreover, we propose that it is possible to laser cool the HgOH molecule by adopting the same technique as that in the diatomic polar molecule, HgF, as shown in [Phys. Rev. A {\bf 99}, 032502 (2019)]. 
\end{abstract}

\maketitle

The electric dipole moment (EDM) of an electron, $d_e$, arises due to the simultaneous violations of parity (P) and time reversal (T) symmetries \cite{1,2}; with the latter implying that the EDM is CP violating, due to CPT theorem \cite{3}. This intrinsic property of the electron is extremely tiny, and its existence has not been confirmed till date; only upper bounds exist. A stringent bound on $d_e$ has significant implications in constraining physics that lies beyond the Standard Model (SM) of particle physics ($\sim$ TeV-PeV energy scales)~\cite{4,5} as well as probing the underlying physics describing the matter-antimatter asymmetry in the universe~ \cite{6,7}. A direct measurement of $d_e$ is almost impractical. Therefore, atoms or molecules are used as a means of probing $d_e$ in table-top experiments. The EDM of an electron in an atom or a molecule can interact with both the intrinsic electric field due to electromagnetic interactions within the system and an external electric field applied for performing the experiment. The magnitude of the intrinsic electric field that an electron with an EDM experiences can be viewed as an effective electric field ($\mathcal{E}_{\mathrm{eff}}$). The effective electric field can be very large in certain heavy polar molecules as compared to  atoms~\cite{9,10,11,Skrip}. Hence, polar molecules are considered as the most promising candidates for inferring the $d_e$ value. In fact, the EDM of an electron is extracted by a combination of the theoretically determined $\mathcal{E}_{\mathrm{eff}}$ with the experimentally measured shift in energy due to $d_e$ interacting with an electric field in a molecule. The quantity, $\mathcal{E}_{\mathrm{eff}}$, can only be obtained by carrying out relativistic many-body calculations~\cite{Schiff, Sandars}. Also, theoretical studies play the crucial roles in identifying suitable molecules, with highly enhanced $\mathcal{E}_{\mathrm{eff}}$, for EDM experiments. 

The last one and a half decades have witnessed tremendous growth in  experimental and theoretical fronts in the search for EDM using diatomic polar molecules. This is evident from the sheer number of ongoing experiments, such as ThO \cite{12}, HfF$^+$ \cite{13}, YbF \cite{14}, BaF \cite{15,15prime}, etc. Among these, the most accurate limit comes from ThO, with $\arrowvert d_e \arrowvert \leq 1.1\times  10^{-29}$ e-cm. On the theoretical side, many interesting proposals have been put forth, identifying new polar molecules and molecular ions, including PtH$^+$, and HfH$^+$~\cite{MandB1}, YbRb, YbCs, YbSr$^+$, and YbBa$^+$~\cite{MandB}, WC~\cite{MandB2}, RaF~\cite{RaF}, TaN~\cite{TaN},  HgX (X=F, Cl, Br, and I)~\cite{10}, HgA (A=Li, Na, and K)~\cite{HgA}, and RaH~\cite{RaH}. The diatomic HgX polar molecules show tremendous enhancement in their effective electric fields. Nevertheless, most of the diatomic candidates offer only marginal advantages (typically one order improvement in sensitivity) on either the experimental or theoretical aspects, as compared to current experimental systems. In other words, there are limitations in the EDM experiments using diatomic molecules to improve measurements of $d_e$ beyond one to two orders. 

The recent proposal to search for EDM using triatomic molecules, by Kozyvrev and Hutzler(~\cite{16} and references therein), opened new avenues in this domain of research. A triatomic's bent molecular structure could offer crucial advantages besides other favourable benefits of the diatomic polar molecules. As a proof-of-principle, the triatomic YbOH was considered as a suitable candidate in the above proposal. This molecule has the advantage of possessing internal co-magnetometer states (due to closely-spaced opposite-parity doublets), which aids in the rejection of systematics. The molecule is also laser-coolable, due to  electronic structure of Yb being very similar to that of alkaline-earth atoms. A large number of these highly polarizable molecules, prepared in the low-lying (010) vibrational state, and trapped in an optical lattice, promises a sensitivity that could exceed that of ThO by four orders of magnitude; {\it albeit} $\mathcal{E}_{\mathrm{eff}}$ of YbOH is smaller by almost a factor of three as compared to ThO \cite{YbOH1,YbOH2,YbOH3,LVS}. In the previous line, (010) means that the vibrational quantum numbers $n_1 = n_3 = 0$, while $n_2 = 1$, where $n_1$, $n_2$, and $n_3$ denote symmetric stretch, bend, and asymmetric stretch, for a triatomic molecule~\cite{16}. 

In this work, we demonstrate that HgOH could be an even more promising polar triatomic molecule than YbOH for electron EDM searches. We performed accurate calculations of the potential energy curve (PEC), and obtained $\mathcal{E}_{\mathrm{eff}}$ and the permanent electric dipole moment (PDM) of HgOH in its ground state, assuming linear geometry, using a relativistic coupled-cluster (RCC) theory. We find that these two quantities are larger than those of YbOH. We also discuss laser-cooling aspects of HgOH by drawing an analogy of its structure with HgF \cite{HgFlc}. In addition to these factors, one can take advantage of favourable features, such as internal co-magnetometer states, of a triatomic molecule, as discussed in \cite{16}, to polarize it fully and reduce systematics due to its internal co-magnetometer states in a low-lying bent vibrational (010) mode of the ground state. 

\begin{figure}[t]
    \centering
    \includegraphics[width=8.5cm,height=6.5cm]{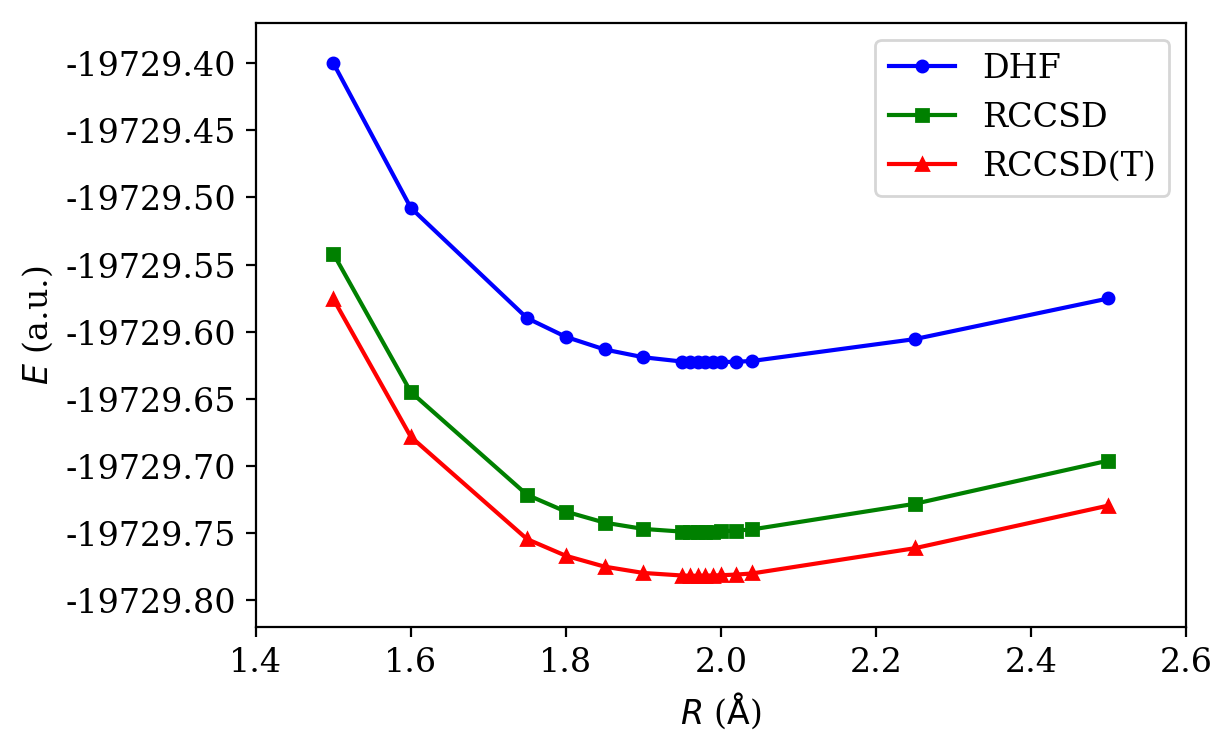}
    \caption{The ground electronic state potential energy ($E$) curve of HgOH in the DHF, RCCSD, and RCCSD(T) approximations, with a double-zeta basis. In the plot, $R$ is the Hg-O bond length. The equilibrium bond length is found to be 1.97 Angstroms from all three methods. The DHF energy has been shifted by $-2.9$ a.u., in order to make it easier to compare the three curves. }
    \label{fig:fig1}
\end{figure}

The electron EDM interaction Hamiltonian in an atomic or molecular system is given by \cite{Lindroth,24}
\begin{equation}
H_{EDM}= 2 icd_e \sum_{i=1}^{N_e}\beta\gamma_5p_i^2,\label{1}
\end{equation}
where $c$ is the speed of light, $\gamma_5$ is a Dirac matrix, $p_i$ is the momentum operator corresponding to the $i^{th}$ electron and $N_e$ is the number of electrons in the system.
The first-order shift in energy level of the molecule, to be measured, due to the above Hamiltonian can be expressed as
\begin{eqnarray}
\Delta E = \frac{\langle\Psi\arrowvert H_{EDM}\arrowvert\Psi\rangle}{\langle\Psi\arrowvert \Psi\rangle}
&=& d_e \frac{\langle\Psi\arrowvert\sum_{i=1}^{N_e} 2 ic \beta\gamma_5p_i^2 \arrowvert\Psi\rangle}{\langle\Psi\arrowvert \Psi\rangle} \nonumber\\
 &=& -d_e\mathcal{E}_{\mathrm{eff}}, 
\end{eqnarray} 
where $\arrowvert\Psi\rangle$ is the wave function due to the electromagnetic interaction. 

In the RCC theory, the wave function of the ground state of a molecule can be expressed as
\begin{equation}
\arrowvert\Psi\rangle=e^T\arrowvert\Phi_0\rangle,
\end{equation}
where $\arrowvert\Phi_0\rangle$ is the Dirac-Hartree-Fock (DHF) wave function and $T=T_1+T_2+\cdots+T_{N_e}$ operator is responsible for hole-particle excitations, with subscript $n=1,2,3, \cdots$ denoting the level of excitation. 

We approximate RCC theory with single and double excitations (RCCSD method) by taking $T\approx T_1 + T_2$ and use the resulting wave function for evaluating $\mathcal{E}_{\mathrm{eff}}$ and PDM ($\mu$), while for the PEC, important contributions from $T_3$ are also included through perturbative approach to the wave function (RCCSD(T) method). We use the Dirac-Coulomb Hamiltonian, e.g. see Ref.~\cite{9}, and electrons from all the core orbitals are excited in our calculations. However, we consider only the virtual orbitals below 1000 atomic units (a.u.) in the single particle orbital energies. 

The expectation value of an operator $O$ can be evaluated in the RCC theory as \cite{Cizek} 
\begin{eqnarray}
\langle O \rangle &=&  \langle\Phi_0\arrowvert e^{T \dagger} O e^T\arrowvert\Phi_0\rangle_l ,
\label{eqedm}
\end{eqnarray}
where subscript, `$l$', means that each of the terms is linked. In our calculations, we have considered about 300 leading-order terms in the diagrammatic approach from this expression, specifically those containing up to quadratic powers in $T$ and $T^{\dagger}$ operators. We obtain $\mathcal{E}_{\mathrm{eff}}$ and $\mu$ values by considering $O \equiv H_{EDM}$ and $O \equiv \sum_{i=1}^{N_{e}} r_i+ \sum_{A=1}^{N_M}Z_Ar_A$, respectively, in the above expression with $N_M$ representing the number of nuclei and $Z_A$ the atomic number of the $A^{th}$ atom in the molecule. 

We first investigated the equilibrium bond length ($R_e$) of the Hg-O bond before calculating $\mathcal{E}_{\mathrm{eff}}$ and $\mu$. We employed the DHF, RCCSD and RCCSD(T) methods using the DIRAC16 program \cite{Dirac16}. We used Dirac-Coulomb Hamiltonian for this purpose. The O-H bond length was chosen as $0.922\si\angstrom$, based on the analysis for YbOH \cite{YbOH3}. This is a reasonable approximation, as we had shown that the properties of interest are  sensitive to the bond between the heavy atom and O, while they change little with the O-H bond length~\cite{YbOH3}. To pin-point the minima, we chose carefully the grid points in the plot, in a non-uniform fashion, so that they were clustered more around the equilibrium bond length. This shows there are large differences between the results from the DHF and RCC theory. We infer the equilibrium bond length of the Hg-O for the ground state of HgOH to be $1.97\si\angstrom$ from the minima of the plot representing the RCCSD(T) results as shown in Fig. \ref{fig:fig1}. Using this bond length, we obtained the $\mathcal{E}_{\mathrm{eff}}$ and $\mu$ values of HgOH in the RCCSD method. We took Dyall's \cite{20} double-zeta (DZ), triple-zeta (TZ) and quadruple-zeta (QZ) basis sets to verify consistency in these results. We used UTChem \cite{21} for carrying out DHF calculations and for atomic orbital (AO) to molecular orbital (MO) integral transformation, and Dirac08 package \cite{22} for the RCCSD calculations. At the AO to MO transformation level, we imposed a $1000$ a.u. energy cut-off for virtuals. We also reported $\mathcal{E}_{\mathrm{eff}}$ and $\mu$ values of the molecule using the DZ basis, with inclusion of all the virtual orbitals (denoted as DZ$^*$), and showed that the values do not differ significantly by including virtuals with orbital energies beyond 1000 a.u.. Our results from different basis sets are presented in Table \ref{tab1} from the DHF and RCCSD methods. As seen, the values for $\mathcal{E}_{\mathrm{eff}}$ change slightly with the basis size. Differences between the DHF and RCCSD results are very small. This implies that it is not necessary to employ a more sophisticated method for calculating $\mathcal{E}_{\mathrm{eff}}$ at this stage. However, we find significant changes in the $\mu$ value with different set of basis and from DHF value to the RCCSD value. Our RCC calculations give $\mu=1.44$ D in a QZ basis, which is slightly larger than $\mu=1.1$ D of YbOH \cite{YbOH3}. We recommend values for both $\mathcal{E}_{\mathrm{eff}}$ and $\mu$ with uncertainties by analyzing change in results due to use of different basis functions and from the DHF values to the RCCSD results. Nevertheless, this level of accuracy is sufficient for carrying out present analysis. It is possible to improve these results further by considering a better approach in the RCC theory framework when the actual EDM experiment comes to fruition. We also compare our $\mathcal{E}_{\mathrm{eff}}$ and $\mu$ of HgOH with the corresponding quantities of ThO and YbOH in the above table.  It is evident from this table that  $\mathcal{E}_{\mathrm{eff}}$ of HgOH is about 1.3 and 4.3 times larger than those of ThO and YbOH , respectively, implying that HgOH is a promising candidate for an EDM search experiment, in general, and as a triatomic molecule, in particular. 

\begin{table}[t]
    \centering
        \caption{$\mathcal{E}_{\mathrm{eff}}$ (in units of GV/cm) obtained using different basis sets. Here, the basis sets with a `*' denotes calculated values without imposing any energy cut-off, while those without it are computed at 1000 a.u. cut-off.}
    \begin{tabular}{lcccc}
    \hline \hline \\
    Basis type  & \multicolumn{2}{c}{$\mathcal{E}_{\mathrm{eff}}$} & \multicolumn{2}{c}{$\mu$} \\
    \cline{2-3} \cline{4-5} \\
         & DHF & RCC & DHF & RCC \\
        \hline \\
        DZ$^*$ & 107.93 & 108.90 & 1.53 &1.00 \\
        DZ & 107.93 & 107.85  & 1.53 &1.00 \\
        TZ & 107.30 & 103.41  & 1.45 &1.44 \\
        QZ & 107.26 & 102.85 & 1.83& 1.44 \\
        & & \\
        Recommended & \multicolumn{2}{c}{103(5)} & \multicolumn{2}{c}{1.44(10)} \\
        \hline \\
         YbOH & \multicolumn{2}{c}{23.80} &  \multicolumn{2}{c}{1.1} \cite{YbOH3} \\
         ThO & \multicolumn{2}{c}{79.9} & \multicolumn{2}{c}{4.24}\cite{LVS} \\
          \hline \hline
    \end{tabular}
    \label{tab1}
\end{table}

As seen earlier, the PDM of HgOH is reasonably large. This can ensure that a low external electric field is sufficient to polarize the molecule, thereby reducing the associated systematics. Another important criterion that serves as guiding light to support a molecule as a favourable candidate for EDM searches is its response to signal-to-noise ratio; i.e. its statistical sensitivity. The statistical sensitivity for the measurement of $d_e$ due to EDM can be estimated by~\cite{18} 
\begin{eqnarray}
\delta d_e^{stat} \sim \frac{1}{\sqrt{NT\tau}\mathcal{E}_{\mathrm{eff}} \eta},\label{7}
\end{eqnarray}
where $N$ is the number of molecules produced in the experiment, $T$ is the total integration time, $\tau$ is the coherence time, and $\eta$ is the polarization factor. The $\eta$ value is directly proportional to $\mu$  of the molecule as well as the external electric field. 

In order to take the advantage of the internal co-magnetometer states of HgOH, we propose that the experiment should be performed in one of the low-lying vibrational bent levels of the ground state, specifically the (010) level, similar to YbOH. In such a case, $\mathcal{E}_{\mathrm{eff}}$ would also depend on the range of available bending angles in that state. Since this range is not known yet for this molecule, the calculations are performed using linear geometry. However, we strongly expect that there may not be significant deviation in the $\mathcal{E}_{\mathrm{eff}}$ value for the (010) level. Therefore, we do not expect the statistical sensitivity to change significantly for a proposed experiment in the (010) bent geometry. To estimate $\delta d_e^{stat}$, we use $\mathcal{E}_{\mathrm{eff}} = 102.85$ GV/cm (that of the linear geometry itself) and $N \sim 10^5$ molecules, total integration time as $10^7$ s, polarization factor as one, and coherence time to be 1 s. This corresponds to statistical sensitivity of $2.015 \times 10^{-32}$ e-cm, which is still comparable to the sensitivity of YbOH whose trapping time is 10 s. On other words, we can anticipate better sensitivity in the HgOH EDM experiment if the above conservative choice of parameters are improved. The choice of number for molecules was decided based on a recent experiment that examined the feasibility of laser cooling the HgF molecule ~\cite{HgFlc}. The integration time that we chose is typical in an electron EDM experiment. Similarly, a lower and conservative value of coherence time was decided keeping in mind its corresponding value of 10 s in the proposal for YbOH experiment \cite{16}. 

\begin{figure}[t]
    \centering
    \includegraphics[width=8cm,height=5.2cm]{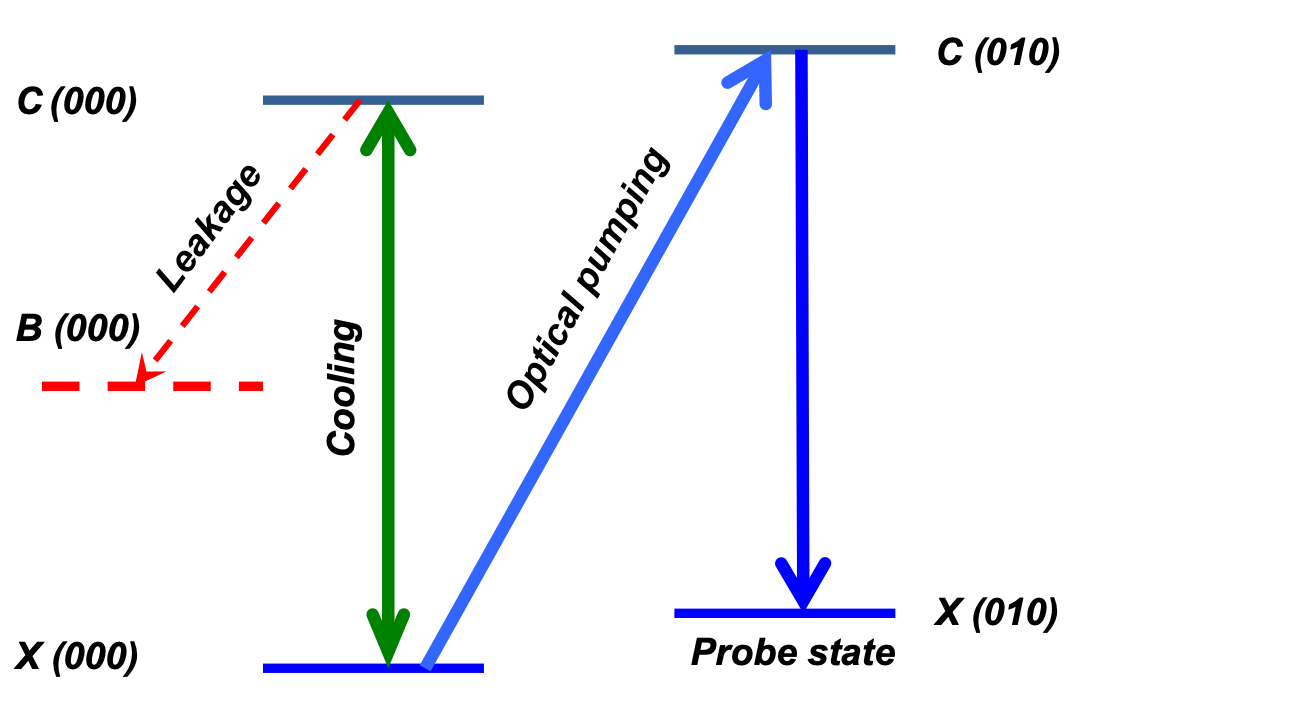}
    \caption{Schematic diagram for laser cooling transition in HgOH, from $X (000)$ to $C(000)$ states. The probe state, which is the vibrational $X (010)$ level, to measure EDM is shown in blue. The dotted $C \rightarrow B$ transition, shown in red, is undesired in this experiment. It is anticipated that the wavelengths of the corresponding transitions will be almost similar to those of HgF. }
    \label{fig:fig3}
\end{figure}

Here, we briefly discuss about the feasibility for laser cooling of HgOH molecule for an electron EDM measurement. It is important to form a closed optical cycle between two electronic states for laser cooling of a molecule. These states need to offer highly diagonal Franck-Condon (FC) matrices~\cite{Bergertri}. Since comprehensive electronic structure calculations of HgOH are beyond the scope of this work, we demonstrate it by arguing that the functional group OH can be treated as a pseudo-halogen, whose chemical properties resemble those of halogens~\cite{Bergertri}. Thus, the electronic structure of HgOH would be almost similar to HgF, where one of the two electrons in the $6s^2$ orbital in Hg atom can form a $\sigma$ bond with the functional group OH, and the other one can occupy the  anti-bonding molecular orbital. The anti-bonding orbital is centered mainly on Hg atom with their center-of-charge shifted away from the bonding region \cite{expt1}. As a result, it can give rise to parallel Born-Oppenheimer potential energy curves for the electronic ground state and an excited state. This can also offer quasi-diagonality to the FC matrix \cite{expt2}. Analogous to the laser cooling technique of HgF \cite{HgFlc}, only a few lasers to re-pump the vibrationally excited states back to the $C$ state are needed to complete the almost closed optical cycling between the $X$ and $C$ states. A leak to the $B$ state, at an estimated rate of $10^{-4}$, will not be able to limit the laser cooling process significantly, but the optical cycling may become more complex~\cite{HgFlc}. The spin-rotation coupling in HgOH is expected to be similar to the case in HgF, however, the hyperfine splitting should be much smaller due to the only hyperfine coupling coming from the far away H atom. One can still employ the microwave remixing method to address all the necessary levels for laser cooling, as proposed for HgF molecules~\cite{HgFlc}. One may encounter the problem of having to scatter more photons during the laser cooling process, owing to heavier mass of HgOH compared to YbOH triatomic molecule. This drawback can be partially offset by large momentum kick from the UV photons emitted from the $X \rightarrow C$ laser cooling transition. The EDM measurement in the triatomic HgOH molecule demands an appropriate pathway for populating the (010) vibrationally excited level from the lower (000) level. Similar to the case in YbOH~\cite{16}, this pathway can be realized by the optical pumping via $X(000) \rightarrow C(010)$, which is not forbidden due to spin-orbit coupling, making it strong enough to drive with a laser. The bending mode in the $C$ state should decay strongly to the bending mode in the $X$ state. Fig. \ref{fig:fig3} gives a schematic of the transitions involved in laser cooling HgOH. We assume that in the figure, the transition wavelengths from $X$ to $C$ states would be close to those of HgF~\cite{HgFlc}, akin to the observed similarities in the $X$ to $A$ transition wavelengths between YbF~\cite{Y1} and YbOH~\cite{Y2}, as well as between CaF~\cite{C1} and CaOH~\cite{C2}. 

In Conclusion, we investigated the effective electric field in HgOH and found that its enhancement is very large as compared to the other triatomic molecule, YbOH, which has been proposed recently for measuring EDM. We also determined its PEC and PDM. The PDM value is seen to be sufficiently large to be able to polarize the HgOH molecule with a fairly low applied electric field in the laboratory. A proof-of-principle for cooling the HgOH triatomic molecule for conducting the experiment is outlined. 

\section{Acknowledgements}

We would like to thank A. C. Vutha, N. R. Hutzler and Z. C. Yan for insightful discussions. All the computations were performed on the Vikram-100 super-computing cluster, PRL, Ahmedabad and we wish to thank Mr. Jigar Raval for his continuous support in successfully running our programs.

\end{document}